\documentclass[12pt]{article}
\usepackage{epsfig}

\newcommand{\eq}{\begin{equation}}
\newcommand{\eqx}{\end{equation}}
\newcommand{\eqn}{\begin{eqnarray}}
\newcommand{\eqnx}{\end{eqnarray}}
\newcommand{\f}[2]{\frac{#1}{#2}}

\newcommand{\qb}{\bar{q}}

\newcommand{\dl}{\delta}
\newcommand{\sg}{\sigma}
\newcommand{\kap}{\kappa}
\newcommand{\xp}{{x_+}}
\newcommand{\xm}{{x_-}}
\newcommand{\qqqq}{\quad\quad\quad\quad}
\newcommand{\LL}{{\cal L}}

\newcommand{\sqpmi}{\sqrt{\xm_i\xp_i}}
\newcommand{\sqpmf}{\sqrt{\xm_f\xp_f}}
\newcommand{\sqf}{\sqrt{1+\f{1}{p^2\xp_f\xm_f}}}
\newcommand{\sqi}{\sqrt{1+\f{1}{p^2\xp_i\xm_i}}}

\newcommand{\cor}[1]{\left\langle{#1}\right\rangle}

\title{The Lund Model at Nonzero Impact Parameter\footnote{Dedicated
to the memory of Bo Andersson.}}

\author{Romuald A. Janik $^{a}$, Robi Peschanski $^b$\footnote{
e-mail: {\tt ufrjanik@if.uj.edu.pl}, {\tt pesch@spht.saclay.cea.fr}}\\ \\
$^a$ \small Jagellonian University,\\
\small Reymonta 4, 30-059 Krakow, Poland.\\
$^b$ \small CEA/DSM/SPhT Saclay\\
\small Unit\'{e} de Recherche associ{\'e}e au CNRS\\
\small CEA-Saclay, F-91191 Gif/Yvette Cedex, France.}

\begin{document}

\maketitle

\begin{abstract}
We extend the formulation of the longitudinal 1+1 dimensional 
Lund model to nonzero
impact parameter using the minimal area assumption. Complete formulae
for the string breaking probability and the momenta of the produced
mesons are derived using the string worldsheet {\em Minkowskian
helicoid} geometry. For strings stretched  
into the transverse dimension, we find probability distribution with
slope linear in $m_T$ similar to the statistical models but without
any thermalization assumptions.  
\end{abstract}

\section{Introduction}

One phenomenological ingredient of models for hadronic reactions is the
description of the hadronization process  when the (hard)
partons produced at the interaction point get dressed to form the observed
hadronic particles which are registered in the detector. 
This is a theoretically difficult problem due to its nonperturbative
character, however phenomenological models are quite successful in the
description of data.
Models which describe this behaviour are essential for the comparision of QCD
predictions for partonic processes with the experimental data.

One of the most successful hadronization models is the Lund model
\cite{LUND,ARTRU} based on the effective string picture of QCD at
strong coupling. The conventional Lund model\footnote{By the
`conventional' Lund model we mean the original 1+1 dimensional
formulation not containing the extension to gluonic strings
\cite{RECENT}.} gives a description of
the break-up of a colour string into produced hadronic states. It is
directly applicable 
in the context of processes where the initial $q\qb$ pair is produced
at a single point, the partons then fly away from each other,
a colour string is stretched between them, which eventually breaks up
producing hadrons. A fundamental property of the confining strings is
the so-called area law \cite{WILSON}, from which the probability
formula in the Lund model may be advocated.

The model is basically 1+1
dimensional,  transverse momenta of the generated mesons are taken
into account by
substituting the masses $m^2$ by $m_T^2=m^2+k_\perp^2$. 
In particular
the transverse momenta are considered to be in principle randomly
distributed, decorrelated along the chain of mesonic
emissions and also uncorrelated with rapidity.   

The aim of this paper is to show that the whole 1+1 dimensional setup
of the Lund 
fragmentation model based on the area law can be generalized to
the 3+1 case when the 
initial quark and antiquark are also separated by a transverse distance $L$
(impact parameter). These types of colour string configurations may
occur between coloured remnants of a projectile and of
the target. They could also appear locally in the hadronization of a
partonic cascade between neighbouring partons at some transverse
distance. 

Our results are the following. We give the full mathematical
derivation of the probability distribution function for the production
of mesons with their 4-momenta ((\ref{e.probdist}) and formulae
(\ref{e.ahel})-(\ref{e.pl})). This is based on the Minkowskian
helicoid geometry 
which is the relevant string worldsheet for nonzero impact
parameter $L$. The results depend on the key parameter $p=\chi/L$, where
$\chi$ is the rapidity interval. For large $p$ we recover the Lund
model probability distribution, while for small $p$ new features
appear. We obtain a linear slope in $m_T$ and not $m_T^2$ as is
usually assumed in effective string models. This derivation based just
on classical considerations has to be contrasted with similar
predictions based on statistical model assumptions.  

The plan of this paper is as follows. We will first review the
standard 1+1 Lund model along the lines of \cite{EXACT}. 
Then in section 3 we will introduce the Minkowskian helicoid minimal
surface which is crucial for the construction of the generalization to
nonzero impact parameter. In section 4 we will show how the basic
formulae of the Lund model get modified in the new setting. In
section 5 we will analyze the limiting cases of small and large $p$. 
In section 6 we close the paper with considerations
of phenomenological interest.

\section{1+1 dimensional Lund model}

Suppose that a highly energetic $q\qb$ pair is produced at the point
$x^\mu=0$. These particles move along straight
lines
\eq
x_0 = \f{v_0}{2} t_0    \qqqq \bar{x}_0 = -\f{v_0}{2} \bar{t}_0 
\eqx
with relative rapidity
\eq
\chi=\f{1}{2} \log \f{1+v_0}{1-v_0} \ .
\eqx
According to the effective string
picture of QCD, a colour string is formed between them.
Since by assumption we are in the confining regime, the string lies in
the most energetically favourable  configuration i.e. it forms
the minimal surface joining the two lines (see fig.~1a). This surface
is of course flat in this case. This follows from the Nambu-Goto
action for a confining string worldsheet $S_{NG}=\kap \cdot Area$.

The process of hadronization is described by the break up of this
string (see fig. 1b) forming a pair of massless quarks, each of which
moves along the light cone directions and which get converted into
hadrons (at the upper tips in fig. 1b). The probability of such a
breakup configuration is given by the area law:
\eq
Probability \sim \exp \left(-b \cdot Area \right)
\eqx   
where $Area$ is the area spanned by the string and $b$ is a phenomenological
parameter which is related to the break-up probability of the string
per unit area \cite{ARTRU}.

\begin{figure}[ht]
\centerline{%
\epsfysize=5cm 
\epsfbox{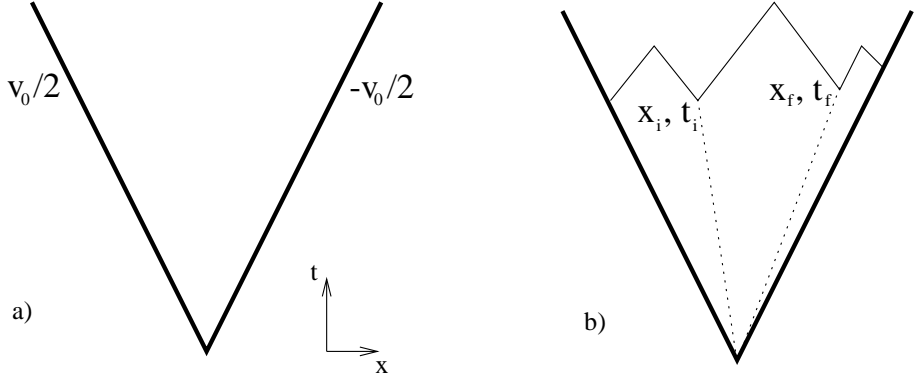}}
\caption{{\em Configuration with zero impact parameter.} a) Initial $q\qb$
configuration; b) a sample configuration of the confining string
with break-up points (the dotted lines divide the string area into
diamond-like sectors -- see text).}
\end{figure}

The area can be calculated as a sum of the areas of the
diamond-shaped sectors (like the one in fig. 1b between the dotted lines). 
Denoting the coordinates of the adjacent points of breakup by
$(t_i,x_i)$ and $(t_f,x_f)$ one obtains the formula
\eq
\label{e.alund}
Area_{if}=\f{1}{4}(2\xm_i\xp_f-\xp_f\xm_f-\xp_i\xm_i)
\eqx
where
\eq
x_{\pm}=t \pm x \ .
\eqx 

At this stage the essential probabilistic set-up of the Lund model is
complete. 

The second ingredient in the Lund model is a prescription how to
associate the produced physical mesons to the string break-up
configuration characterized by the break-up points $(t_k,x_k)$. 

Let the string worldsheet be parameterized by the coordinates $\tau$
and $\sg$. Then the Nambu-Goto lagrangian
\eq
\LL= \kap \sqrt{-\left(h_{\tau\tau}h_{\sg\sg}-h_{\tau\sg}^2\right)} 
\eqx
is expressed in terms of the induced metric
\eq
h_{ab}=\partial_a x^\mu \partial_b x_\mu  \ .
\eqx

The 4-momentum of the produced meson formed between the adjacent points of
breakup $(t_i,x_i)$ and $(t_f,x_f)$ follows from the formula for the
momentum of a (classical) string stretched between these points \cite{BOOK}:
\eq
\label{e.pmu}
P^\mu=\int_C \f{\partial \LL}{\partial \dot{x}^\mu} d\sg -
\f{\partial \LL}{\partial {x'}^\mu} d\tau
\eqx
where $C$ is (any) curve
joining the breakup points.
For flat ($h_{ab}=\eta_{ab}$) or more generally {\em conformally} flat
coordinates ($h_{ab}=e^{\Phi(\sg,\tau)} \eta_{ab}$) this simplifies to 
\eq
\label{e.psimp}
P^\mu=\kap \int_C \left( \dot{x}^\mu d\sg+x'^\mu d\tau \right) \ .
\eqx

Applying (\ref{e.pmu}) to the flat minimal surface shown in fig 1b yields
\eqn
P^0 &=& \kap(x_f-x_i) \nonumber\\
P^1 &=& \kap(t_f-t_i) \nonumber\\
P_\perp &=& 0  \ .
\eqnx
The final ingredient, which is not derived from the model, is the mass
shell condition for the produced mesons. Thus $\dl(P^2-m^2)$ is put in
where $m^2$ are the physical meson masses.

The resulting probability distribution function \cite{LUND} for the
production of mes\-ons with momenta $\{p_j\}$ is thus:
\eq
\label{e.probdist}
dP_n(\{p_j\},P_{tot}) \propto \prod_{j=1}^n  d^2 p_j\, \dl(p_j^2-m^2_j)\,
\dl\!\left(\sum_jp_j-P_{tot} \right) \exp(-b \cdot Area) 
\eqx 
where
\eq
Area \equiv \sum Area_{if}
\eqx
is the total area of the string worldsheet (see fig.~1b). 

Note that all the above
steps and formulae, apart from the mass-shell condition, follow just from
the geometry of the flat surface between the initial lines, which in
turn is unambigously determined as the minimal surface between the
trajectories of the initial energetic $q\qb$ pair which generates the
initial colour string. 
The point that we would like to emphasize here is that it is possible
to use this property to generalize the model to configurations with
nonzero impact parameter.

\section{The minimal surface geometry: the Mink\-owskian helicoid}

\begin{figure}[ht]
\centerline{%
\epsfysize=5cm 
\epsfbox{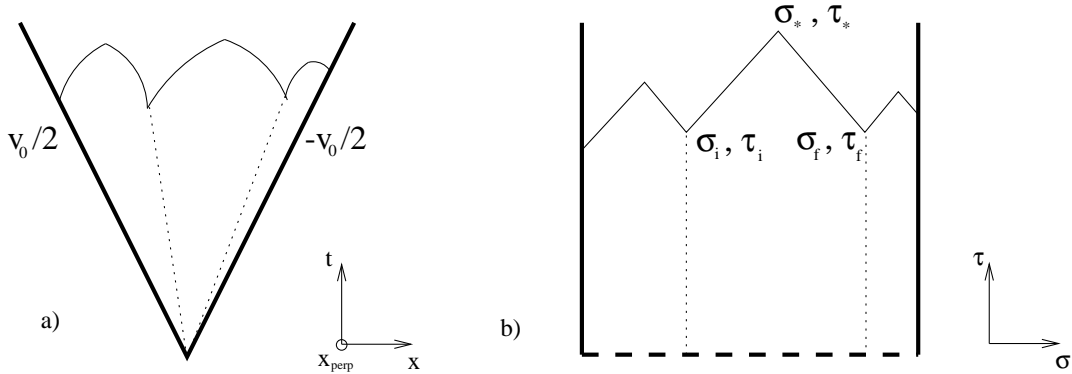}}
\caption{{\em A string breaking on the helicoid.} a) picture projected
onto the longitudinal plane (to be compared with fig. 1b); b) the same
configuration in the $\tau-\sg$ variables.} 
\end{figure}

When the string is formed between a quark and antiquark, moving with
relative rapidity $\chi$, and 
separated by a transverse distance $L$, the string worldsheet will
stretch along the minimal surface formed by the two lines. 
An Euclidean version of this surface has been used in an approach to
study high energy scattering using the AdS/CFT correspondence
\cite{US}. 

The minimal surface here\footnote{The {\em assumption} that the string at $t=0$ 
is a straight segment in impact parameter selects the helicoid minimal surface. 
If
the initial string would be in some excited configuration the analysis
would  possibly lead to different solutions.  Note that an {\it a priori} 
different transverse configuration, going beyond the original Lund model, has 
been studied in \cite{coherent} with a numerical treatment based on the ARIADNE 
2 program \cite{ariadne}.} is a (Minkowskian) M-helicoid which can be
para\-metrized by 
\eqn
t &=& \f{1}{p} \sinh p \tau \cosh p \sg \nonumber\\
x &=& \f{1}{p} \sinh p \tau \sinh p \sg \nonumber\\
x_\perp &=& \sg
\eqnx 
where $p=\chi/L$, $\sg=-L/2 \ldots L/2$ and $\tau$
starting\footnote{Our formalism can be easily extended to other
initial conditions for $\tau.$} at 0.  
The string at $t=0$, when the particles are separated by a transverse
distance equal to the impact parameter, is stretched along a straight
line. 

The geometry of the helicoid is
encoded in the induced conformally flat metric:
\eq
h_{ab}=(\cosh^2 p\tau) \cdot \eta_{ab} \ .
\eqx 
Hence
\eq
ds^2 = \cosh^2 p\tau \cdot (d\tau^2-d\sg^2) \ .
\eqx
Since the produced quarks are assumed to be massless, they follow
light-like trajectories ($ds^2=0$) on the helicoid which are just straight
lines at an angle of $\pm 45^o$ in the $\tau-\sg$ variables. The area
element is
\eq
d{\cal A}=\sqrt{-\det h}=\cosh^2 p\tau \, d\tau d\sg  \ . 
\eqx
 
It is also convenient to look at the helicoid projected onto the
longitudinal $t-x$ plane. Then one can show that
\eq
\label{e.sg}
\f{x}{t}=\tanh p\sg   \qqqq p\sg=\f{1}{2}\log\f{\xp}{\xm} \ .
\eqx
So the lines of constant $\sg$ correspond to fixed longitudinal
velocities\footnote{This is also the definition of the boundaries of
the sectors in fig. 1b in the original zero impact parameter model.}
$v$.  

The $\tau$ coordinate can also be related to the longitudinal
coordinates through
\eq
\sinh p\tau =p \sqrt{\xp \xm}  \ .
\eqx

As we saw in the previous section all ingredients of the Lund model
can be formulated as purely geometric constructions based on the
minimal surface which represents the colour string. We thus have to
redo the calculations of section 2 substituting the helicoid for the
original flat surface.

\section{The Lund model at nonzero impact parameter}

We will now derive the formulae for the break-up probability and
produced momenta for the nonzero impact parameter case which are
enough to define the generalization of the full probability
distribution function (\ref{e.probdist}). The only thing
that changes is the underlying geometry which is no longer flat but
is given by the Minkowskian helicoid. It is convenient to label the
breakup points by the longitudinal coordinates $(t,x)$
(i.e. to look at the helicoid projected onto the $t-x$ plane) and
subdivide the area into same sectors as before (the boundaries being
lines of constant slope $x/t=v=const$, i.e. constant $\sg$). In
fig.~2a we show the breakup in the $t-x$ plane while in fig.~2b we
depict the {\em same} process\footnote{We should emphasize that in contast to 
the break-up points, which have
a definite physical meaning, the subdivision of the string worldsheet
into sectors is to a large extent arbitrary. Our choice here was to
subdivide the area in a way similar to the flat string
case in order to facilitate the comparison with the 1+1 dimensional
Lund model. Of course the final probabilistic formulas contain only
the {\em sum} of all sectors which is uniquely defined in the $\tau -\sg$ 
variables.}.
. 

We calculate first the area of the sector defined by the breaking
up points $(\sg_i,\tau_i)$ and $(\sg_f, \tau_f)$. The coordinates of
the tip of the sector are:
\eq
\sg_*=\f{\tau_f-\tau_i+\sg_f+\sg_i}{2} \qqqq
\tau_*=\f{\tau_f+\tau_i+\sg_f-\sg_i}{2}  \ .
\eqx
The area is then given by
\eq
Area_{if}=\int_{\sg_i}^{\sg_*} d\sg \int_0^{\tau_i+\sg-\sg_i} \cosh^2
p\tau\, d\tau \, +
\int_{\sg_*}^{\sg_f} d\sg \int_0^{\tau_f-\sg+\sg_f} \cosh^2
p\tau\, d\tau
\eqx
Using the integral
\eq
\int_0^s \cosh^2 p\tau\, d\tau=\f{s}{2}+\f{\sinh(2p s)}{4p}
\eqx
the $Area_{if}$ can be calculated in closed form:
\eqn
\label{e.ahel}
\!\!\!\!Area_{if}\!\!&=&\!\!\f{1}{8} \left( (\sg_f-\sg_i)^2 - (\tau_f-\tau_i)^2
+2(\sg_f-\sg_i)(\tau_f+\tau_i) \right) + \nonumber\\
\!\!\!\!&&\!\!\!\!\!\!\!\! \!\!\!\! +\f{1}{8p^2} \left(
2\cosh(p(\sg_f-\sg_i+\tau_f+\tau_i))-\cosh(2p\tau_f)-\cosh(2p\tau_i)
\right)
\eqnx

We now have to compute the momentum carried by the string between the
breakup points. This is identified with the 4-momentum of the produced
meson. Using formula (\ref{e.psimp}) we get
\eqn
\label{e.psgtau}
P^0 &=& \f{\kap}{p} \left( \cosh p\tau_f \sinh p\sg_f-\cosh p\tau_i \sinh
p\sg_i \right) \nonumber \\
P^1 &=& \f{\kap}{p} \left( \cosh p\tau_f \cosh p\sg_f-\cosh p\tau_i \cosh
p\sg_i \right) \nonumber \\
P_\perp &=& \kap (\tau_f-\tau_i) \ .
\eqnx
It is convenient, also for comparison with the zero impact parameter
case to rewrite this in terms of longitudinal coordinates
\eqn
\label{e.pl}
P^0 &=& \kap \left( \sqf x_f-\sqi x_i \right) \nonumber\\
P^1 &=& \kap \left( \sqf t_f-\sqi t_i \right) \nonumber \\
P_\perp &=& \f{\kap}{p} \left(
\log\left[\sqrt{1+p^2\xp_f\xm_f}+p\sqrt{\xp_f\xm_f} \right] -(f \to i) \right)
\ .
\eqnx

Formulae (\ref{e.ahel}), (\ref{e.psgtau}) and equivalently
(\ref{e.pl}) when inserted into (\ref{e.probdist}) completely define
our model and can be used as a basis of a Monte-Carlo simulation.

\section{The limits of small and large aspect ratio $p$}

The key parameter which characterizes the surface geometry is
$p=\chi/L$. The scale of the momenta of the produced mesons is set by
the string tension $\kap$.

\subsubsection*{Small impact parameter}

Let us first verify that when $p\to \infty$ (large rapidity or zero impact
parameter limit) we recover the formulae of the conventional Lund model.
Indeed it is clear from (\ref{e.pl}) that the momenta of
the particle produced go over to the conventional Lund expressions
(note that $P_\perp \sim (\log p)/p \to 0$). In fact we have to impose
the condition that $x_\pm$ of the break-up points are kept
{\em fixed} in the limit. From (\ref{e.sg}) this means that $\sg \to 0$ with
$p\sg$ fixed.

The $Area$ of each sector (\ref{e.ahel}) also reduces to the Lund expression
(\ref{e.alund}). To see this we note that for large $p$ (and keeping
the longitudinal coordinates of the break up points fixed) the
polynomial terms in (\ref{e.ahel}) vanish, the only contribution comes
form the hyperbolic cosines. This can be easily evaluated using
$e^{p\sg}=\sqrt{\xp/\xm}$ and $e^{p\tau} \sim 2p \sqrt{\xp \xm}$ (for
$p \to \infty$). The result is the Lund expression (\ref{e.alund}).

Since we obtained the expression (\ref{e.alund})
where there is no explicit remaining $p$ dependence, our assumption
that the break up longitudinal coordinates do not scale with $p$ but
remain of order 1 is self-consistent. In this way we verified that the
conventional Lund model is a smooth limit of the nonzero impact
parameter generalization.

\subsubsection*{Large impact parameter}

It is especially interesting to study which new phenomenae may arise
in comparision to the original model, and whether these are phenomenologically
important. To answer this question in full one has to perform a Monte
Carlo simulation. Here we will just consider the limit of
large impact parameter (small $p$ limit) where one can expect the
largest deviations from the original Lund model.

Before we make the analysis let us note that the string worldsheet in
the $p \to 0$ limit is a flat string stretched in the {\em transverse}
plane which is still unbroken at $t=0$. Then it will fragment into
mesons which will have nonzero transverse momenta even at the
classical level. 

Let us check how does the area scale when $p \to 0$. If we where
to keep the longitudinal coordinates fixed and of order 1 we would
arrive at a singular area\footnote{One gets then
\[
Area_{x_\pm}=\f{1}{8p^2} \left[ \f{\xm_i\xp_f+\xm_f
\xp_i}{\sqpmi \sqpmf}-2 \right] + \f{1}{32p^2} \left( \log
\f{\xp_f}{\xm_f}-\log \f{\xp_i}{\xm_i} \right)^2 \ .
\]}
(`action') for $p\to 0$. Hence the above assumption in this case is
unjustified. 
The correct choice is in fact that the coordinates $(\sg,\tau)$ remain
of order 1. This is natural physically as the transverse mass 
in the small $p$ limit stays finite and becomes $m_T=\kap \Delta \sg
\equiv \kap(\sg_f-\sg_i)$. The $\tau$ coordinate is also directly linked
to the transverse momentum through $\Delta \tau=P_\perp/\kap$.

Considering now $\sg$ and $\tau$ fixed and starting with the area
formula (\ref{e.ahel}) one gets for $p\to 0$: 
\eq
Area_{if} \to \f 14 \left[\Delta^2\sg-\Delta^2 \tau\right] +\tau^+\ \Delta \sg 
\sim \f{m^2}{4\kap^2}+\tau^+\ \f{m_T}{\kap} ,
\label{e.smallp}
\eqx
where
\eq
\Delta\sg \equiv \sg_f\!-\!\sg_i \ ; \Delta \tau \equiv \tau_f\!-\!\tau_i\ \ ; 
\tau^+\equiv \f{\tau_f\!+\!\tau_i}2 \ .
\eqx

Therefore the break-up probability for the formation of a meson takes
the following simple form
\eq
Probability \propto \exp \left(-\f{b m^2}{4 \kap^2}\right)\cdot   
\exp {\left(- \f{b m_T}{\kap} \tau^+ \right)}\ ,
\label{Prob}
\eqx   
where $\tau^+$ can be related to the mean time value of two
adjacent break-up points in the limit $p\to 0.$ Indeed, one has 
\eq
\sqrt {t_i^2-x_i^2}+ \sqrt {t_f^2-x_f^2} = \f 1p \sinh {p\tau_i} +\f
1p \sinh {p\tau_f} \to \tau_i+\tau_f \ when \ p\to 0\ .
\label{lim}
\eqx

\section{Phenomenological considerations}

Formula (\ref{Prob}) has quite intriguing properties. Let us assume
that the value of $\tau^+$ fluctuates around some
average value $\cor{\tau}$. We emphasize that this assumption should
be checked in a complete Monte-Carlo simulation.
Then the distribution of produced particles is predicted to follow a
simple formula
\eq
\label{e.avg}
\exp \left(-\f{b m^2}{4 \kap^2}\right)\cdot   
\exp {\left(- \f{b m_T}{\kap} \cor{\tau} \right)}
\eqx
with a slope {\em linear} in $m_T$. This is quite unusual in
string based models where generically a behaviour quadratic in $m_T^2$
appears\footnote{In ref. \cite{Bialas} it was suggested that the
phenomenologically preferred behaviour linear in $m_T$ can be obtained
by considering gaussian fluctuations of the string tension.}
\cite{Bialas} in relation with the Schwinger mechanism
\cite{Schwinger}. Here the slope linear $m_T$  arises due to subtle
cancellations between quadratic terms in the helicoid area and thus
appears directly at the classical level.  

It is interesting to compare these properties with those of the generic
statistical model \cite{HAGEDORN} which successfully describes both the
slope in $m_T$ and the abundances of different particle species in
high energy reactions \cite{Beccatini}. In this model the creation of
particles is supposed to come from thermalized clusters of limited center
of mass energy. Our model, at small $p$, leads to similar conclusions
but without invoking the thermalization assumption. Indeed we have checked
that the slope in $m_T$ and the average relative distribution of the
multiplicities of different species can be qualitatively described by
a simple formula of the form (\ref{e.avg}). In order to go further and
to determine the parameters like $\cor{\tau}$ one needs to
perform a complete Monte-Carlo simulation which goes beyond the scope
of this paper.

Let us finally note another possible implication of the string
breaking model based on the helicoid geometry. Since the helicoid is
extended in the transverse direction, transverse momenta get produced
at the classical level on the same footing as longitudinal ones. This
opens up an exciting possibility that correlations between
$p_t$ and rapidity could be naturally incorporated in the model. We
leave a detailed investigation of these issues which should be based
on a Monte Carlo simulation for future work\footnote{We thank the
  referee for suggesting this problem.}.

\bigskip

\noindent{}{\bf Acknowledgments.} A discussion with our deeply missing friend Bo 
Andersson at the Weimar Conference (2001) is to be acknowledged. We thank Gosta 
Gustafson, Gunnar
Ingelman and Torbjorn Sjostrand for discussions.
RJ would like to thank SPhT Saclay for hospitality when this work was
completed. RJ was partially supported by KBN grant~2P03B09622
(2002-2004).


\begin{thebibliography}{99}


\bibitem{LUND} 
B.~Andersson, G.~Gustafson, G.~Ingelman and T.~Sjostrand,
Phys.\ Rept.\  {\bf 97} (1983) 31;
B.~Andersson, G.~Gustafson and B.~Soderberg,
Z.\ Phys.\ C {\bf 20} (1983) 317;

\bibitem{ARTRU}
X.~Artru and G.~Mennessier,
Nucl.\ Phys.\ B {\bf 70} (1974) 93;
X.~Artru,
Phys.\ Rept.\  {\bf 97} (1983) 147.

\bibitem{RECENT}
B.~Andersson, S.~Mohanty and F.~Soderberg,
arXiv:hep-ph/0212122.



\bibitem{WILSON}
K.~G.~Wilson,
Phys.\ Rev.\ D {\bf 10} (1974) 2445.


\bibitem{EXACT}
B.~Andersson and F.~Soderberg,
Eur.\ Phys.\ J.\ C {\bf 16} (2000) 303
[arXiv:hep-ph/9910374].

\bibitem{BOOK} See e.g. formula (3.31) in B.~M.~Barbashov and
V.~V.~Nesterenko, ``Introduction To The Relativistic String Theory,''
Singapore, World Scientific (1990).

\bibitem{US}
R.~A.~Janik and R.~Peschanski,
Nucl.\ Phys.\ B {\bf 586} (2000) 163
[arXiv:hep-th/0003059];
R.~A.~Janik,
Phys.\ Lett.\ B {\bf 500} (2001) 118
[arXiv:hep-th/0010069];
R.~A.~Janik and R.~Peschanski,
Nucl.\ Phys.\ B {\bf 625} (2002) 279
[arXiv:hep-th/0110024].


\bibitem{coherent} 
B.~Andersson, G.~Gustafson, L.~L\"onnblad and U.~Petterson,
Z.\ Phys.\ C {\bf 43} (1989) 625.

\bibitem{ariadne} 
U. Pettersson: LU-TP-88-5; L. Lonnblad, U. Pettersson  LU-TP-88-15.

\bibitem{Bialas}
A.~Bialas,
Phys.\ Lett.\ B {\bf 466} (1999) 301
[arXiv:hep-ph/9909417].

\bibitem{Schwinger}
J.~S.~Schwinger,
Phys.\ Rev.\  {\bf 128} (1962) 2425.

\bibitem{HAGEDORN}
R.~Hagedorn,
Nuovo Cim.\ Suppl.\  {\bf 3} (1965) 147.

\bibitem{Beccatini} 
F.~Becattini,
Z.\ Phys.\ C {\bf 69} (1996) 485;
F.~Becattini and G.~Passaleva,
in high energy collisions,''
Eur.\ Phys.\ J.\ C {\bf 23} (2002) 551
[arXiv:hep-ph/0110312].




\end{thebibliography}
\end{document}